\documentclass[12pt]{article}
\usepackage[a4paper, margin=2cm]{geometry}
\usepackage{natbib}
\usepackage{amsmath}
\usepackage{amsfonts}
\usepackage{fancyvrb}
\usepackage{caption}
\usepackage{pbox}
\usepackage{float}
\usepackage{multirow}
\usepackage{hyperref}
\usepackage{authblk}
\usepackage{tablefootnote}
\usepackage{amsmath}
\usepackage{amsfonts}
\usepackage{enumerate}
\usepackage{mathtools, nccmath}
\usepackage{url} 
\usepackage{xcolor}
\usepackage{setspace}
\usepackage{tabularx}
\usepackage{mathtools}
\usepackage{adjustbox}
\usepackage{multirow}
\usepackage[flushleft]{threeparttable}
\usepackage{listings}
\usepackage{subcaption}
\usepackage[title]{appendix}
\usepackage{float}
\usepackage{amssymb}
\usepackage{enumitem}
\usepackage{bbm}
\usepackage{verbatim}
\usepackage{longtable}
\newcommand{\pkg}[1]{{\normalfont\fontseries{b}\selectfont #1}}
\newcommand{\noop}[1]{}
\let\proglang=\textsf
\let\code=\texttt
\title{On integrating the number of synthetic data sets $m$ into the \textit{a priori} synthesis approach}
\author[$\dagger$]{James Jackson}
\author[$\star$]{Robin Mitra}
\author[$\dagger$]{Brian Francis}
\author[$\ddag$]{Iain Dove}
\affil[$\dagger$]{Lancaster University, Lancaster, UK}
\affil[$\star$]{Cardiff University, Cardiff, UK}
\date{}
\affil[$\ddag$]{Office for National Statistics, Titchfield, UK}
\begin{document}
\maketitle
\begin{abstract}
Until recently, \textit{multiple} synthetic data sets were always released to analysts, to allow valid inferences to be obtained. However, under certain conditions - including when saturated count models are used to synthesize categorical data - single imputation ($m=1$) is sufficient. Nevertheless, increasing $m$ causes utility to improve, but at the expense of higher risk, an example of the risk-utility trade-off. The question, therefore, is: which value of $m$ is optimal with respect to the risk-utility trade-off? Moreover, the paper considers two ways of analysing categorical data sets: as they have a contingency table representation, multiple categorical data sets can be averaged \textit{before} being analysed, as opposed to the usual way of averaging post-analysis. This paper also introduces a pair of metrics, $\tau_3(k,d)$ and $\tau_4(k,d)$, that are suited for assessing disclosure risk in multiple categorical synthetic data sets. Finally, the synthesis methods are demonstrated empirically.
\end{abstract}
\section{Introduction}
When disseminating data relating to individuals, there are always two conflicting targets: maximising utility and minimising disclosure risk. To minimise risk, statistical disclosure control (SDC) methods, which typically involve either suppressing or perturbing certain values, are applied to a data set prior to its release. One such method is the generation of synthetic data sets \cite{Rubin1993,Little1993}, which involves fitting, and then simulating from, models fit to the original data. These methods, while reducing risk, adversely impact the data’s utility resulting in a clear trade-off between risk and utility.  \par This paper focuses on the role of multiple data sets when synthesizing categorical data (that is, data consisting of only categorical variables) at the aggregated level using saturated count models \citep{Jackson2021}. Saturated synthesis models invite a bespoke approach to synthesis.  They allow the synthesizer to generate synthetic data with certain pre-specified properties, thus allowing them to easily tailor the synthesis to suit the data environment \citep{Elliot2018}. For example, if the intention is to release open data, relatively more noise can be applied to the data than if the data are released in a secure environment. While the Poisson log-linear model is often used to model categorical data, for synthesis this is not necessarily an optimal choice, because the synthesizer - that is, the person(s) responsible for synthesizing the data - has no control over the variance and has, therefore, no way to add noise to at-risk records in the data. For this reason, the negative binomial (NBI), a two-parameter count distribution, is much more effective for synthesis. As the NBI distribution{'s variance is not completely determined by the mean} - though the variance is always greater than the mean - the variance \textit{can} be increased accordingly. Nevertheless, there are still restrictions and these are discussed later on. \par Specifically, this paper explores how flexibility can be incorporated into the mechanism through the use of multiple synthetic data sets. In some cases (as explained in Section \ref{sec3}), $m>1$ synthetic data sets must be generated; while in other cases, though it may be sufficient to generate just $m=1$ synthetic data set, the optimal $m$ can still be considered in relation to the risk-utility trade-off: does the improvement in utility sufficiently outweigh the cost in terms of greater risk? This is because, since it reduces simulation error, increasing $m$ leads to greater utility but also, inevitably, greater risk \citep{Reiter2005b,Reiter2009}. More generally, considering $m>1$ introduces another tuning parameter for the synthesizer to set, thereby providing further flexibility. \par This paper is structured as follows: Section \ref{sec2} summaries the $(\sigma, \alpha)$-synthesis mechanism, on which the results in this paper are based; Section \ref{sec3} extends the mechanism to incorporating $m>1$; Section \ref{sec4} introduces the $\tau_3(k,d)$ and $\tau_4(k,d)$ metrics, developed to assess risk in multiple categorical synthetic data sets; Section \ref{sec5} presents an illustrative example; and lastly the Discussion proposes some areas or future research.
 \section{Review of the use of saturated models for synthesis} \label{sec2}
 The discrete nature of categorical data allow it to be expressed as a multi-dimensional contingency table (multi-way table). As a multi-way table, the data consist of a structured set of cell counts $f_1,\hdots,f_K$, which give the frequencies with which each combination of categories is observed. \par Synthetic data sets can then be generated by replacing these observed counts (known henceforth as  ``original counts'') with synthetic counts. This can be achieved through a count model, such a Poisson log-linear model or a Poisson-Gamma model \citep{Graham2007}. \par Data sets which include large categorical variables (large number of categories), such as administrative databases, return high-dimensional tables. It can be time-consuming to fit log-linear models, for example, to such large tables using standard statistical software, such as the \code{glm} function in \proglang{R}. The iterative proportional fitting (IPF) algorithm can alleviate this problem to a certain extent, however this only returns fitted values - not standard errors - and there reaches a point (even larger tables) where IPF becomes infeasible. \par The use of saturated models, therefore, is beneficial, because obviously no actual fitting is needed to obtain the fitted counts. There are two distinct modelling methods for contingency tables: multinomial models and count models. The multinomial ensures that the total number of individuals in the original data $n$, is equal to the total number of individuals in the synthetic data $n_\text{syn}$. The \code{syn.catall} function in the \proglang{R} package \pkg{synthpop} \citep{Nowok2016} can be used to generate synthetic data via a saturated multinomial model. \par The $(\sigma, \alpha)$-synthesis mechanism \citep{Jackson2021}, on the other hand, uses saturated \textit{count} models for synthesis; specifically, either a saturated negative binomial (NBI) model or a saturated Poisson-inverse Gaussian (PIG) \citep{Rigby2019} model. In this paper, only the NBI has been considered, to emphasize the tuning capability of $m$. Besides, the results from using either the NBI or PIG distributions are broadly similar, as they share the same variance function. \par When $m=1$, the $(\sigma, \alpha)$-synthesis mechanism has two parameters which are set by the synthesizer. The first, denoted by $\sigma$, originates from using a two-parameter count distribution (the NBI or the PIG), parameterised in such a way that $\sigma$ controls the variance. The value of $\sigma$ can be adjusted to inflate - or deflate - the variability in the synthetic cell counts, thus increasing - or decreasing - the expected divergence from the original counts. More noise is required to sensitive cells - usually small cell counts, which correspond to individuals who have a unique (or near-unique) set of observations - to mask that these are, indeed, unique. \par The mechanism’s second parameter, denoted by $\alpha$, relates to the size of the pseudocount - in practice, this is not actually a count but a very small positive number such as 0.01 - which is added to zero cell counts (zero cells) in the original data. This assigns a non-zero probability that a zero cell is synthesized to a non-zero. The pseudocount $\alpha$ is only applied to non-structural zero cells (known as random or sampling zeros), which are zero cells for which a non-zero count \textit{could }have been observed. Throughout this paper it has been assumed, for brevity, that $\alpha=0$. \par Given an original count $f_i=N$ ($k=1,\hdots,K$), the corresponding synthetic count $\bar{f}_i^\text{syn}$ is modelled as:
\begin{align}
f^\text{syn}_i \mid f_i=N_1, \sigma &\sim \text{NBI}(N_1, \sigma), \quad \text{that is,} \quad \nonumber  \\
p(f^\text{syn}_i=N_2 \mid f_i=N_1, \sigma) &=  \frac{\Gamma(N_2+{1}/{\sigma})}{\Gamma(N_2+1) \cdot \Gamma({1}/{\sigma})} \cdot \bigg(\frac{\sigma N_1}{1+\sigma N_1}\bigg)^{N_2} \cdot \bigg(\frac{1}{1+\sigma N_1}\bigg)^{1/\sigma}.  \nonumber
\end{align}
The shape parameter $\sigma$ is set by the synthesizer to adjust the variance of the model. The NBI distribution's variance function is $\text{Var}(f^\text{syn}_i  \mid f_i=N,\sigma)= N + {\sigma}{N^2}$, which demonstrates how the parameter $\sigma$ controls the variance: for example, when $N=1$ the variance is $1+\sigma$, hence a linear relationship. \par
 Using a saturated count model has certain advantages for the unique situation that is modelling for the purpose of synthesis. Firstly, it guarantees the preservation of relationships between variables, as no assumptions are made as to which interactions exist. Secondly, the method scales equally well to large data sets, as no model fitting is required - the model’s fitted counts are just the observed counts. Finally, as the fitted counts are just equal to the observed counts, it allows expected properties of the synthetic data to be determined \textit{a priori} (that is, prior to synthesis). The uncertainty from modelling is, in effect, drained away, and instead uncertainty is injected where it is most needed: to add noise to sensitive cells in the original data. 
\subsection{The $\tau$ metrics}
The following $\tau$ metrics \citep{Jackson2021}, give a basic quantification of risk (and utility):
\begin{align*}
\tau_1(k)&=p(f^{\text{syn}}=k) & \tau_3(k)&=p(f^{\text{syn}}=k | f=k) \\ \tau_2(k)&=p(f=k) & \tau_4(k)&=p(f=k | f^{\text{syn}}=k),
\end{align*}
where $f$ and $f^\text{syn}$ are arbitrary cell counts in the original and synthetic data, respectively. \par Firstly, $\tau_2(k)$ is the empirical proportion of cells in the \textit{original data} with a count of $k$; the value of $\tau_2(k)$ is obtained by dividing the number of cells with a count of $k$ over the total number of cells $K$. Similarly, $\tau_1(k)$ is the proportion of cells in the \textit{synthetic data} with a count of $k$. \par The metric $\tau_3(k)$ is the probability that a cell of size $k$ in the original data is synthesized to $k$; and $\tau_4(k)$ is the probability that a cell of size $k$ in the synthetic data originated from a cell of size $k$. The metrics $\tau_3(k)$ and $\tau_4(k)$ are the most associated with risk; particularly $\tau_3(1)$ and $\tau_4(1)$, since these relate to uniques and there is not the same `safety in numbers' compared with larger cell counts. When, for example, $\tau_4(1)$ is high (close to 1), then it is possible to identify, with near certainty, some uniques in the original data from the synthetic data. \par When saturated models are used, the expected values of these $\tau$ metrics can be found analytically, as functions of the parameters available to the synthesizer ($\sigma$, $\alpha$ and, as later described, $m$). Hence the synthesizer is aware, \textit{a priori}, of the noise required to achieve a specified $\tau_3(1)$ or $\tau_4(1)$ value. 
\section{The role of $m$ as a tuning parameter} \label{sec3}
When generating fully synthetic data in the traditional sense \citep{Raghunathan2003}, multiple synthetic data sets \textit{must} be generated in order to obtain valid inferences, because the synthesis utilises the Bayesian framework developed for the multiple imputation of missing data \citep{Rubin1987}. Thus multiple data sets are needed to properly quantify the uncertainty arising from imputation, when subsequently analysing the synthetic data sets. However, when values are just replaced rather than imputed, as is done in the $(\sigma, \alpha)$-synthesis mechanism, multiple data sets are unnecessary. So while $m>1$ data sets are not fundamental to obtaining \textit{valid} inferences, the \textit{quality} of inferences, for example, the width of confidence intervals, can still be improved upon by increasing $m$. Thus, prior to releasing synthetic data, the synthesizer can tweak $m$ to find an optimal balance between risk and utility. The motivation for keeping $m$ small is two-fold: firstly, larger $m$ typically results in heightened disclosure risk \citep{Reiter2005b,Reiter2009}; and secondly, it relieves the pressure on storage demands, which is especially relevant when synthesizing large administrative databases. \par Although there is obviously a greater computational burden when generating multiple data sets, the effect on central processing unit (CPU) time can be substantially reduced by making use of parallel processing. As the $m>1$ data sets are independently generated, this is a classic example of a task that can be undertaken in this way. Parallel processing utilises multi-core CPUs; for example, if the task is carried out on a CPU with four cores rather than one, then the CPU time would be roughly four times faster. Besides, for the NBI, the CPU time is quick anyway; the example synthesis presented in Section \ref{sec5} only took approximately 0.3 seconds for $m=1$ with $K=3.5 \times 10^6$ on a typical laptop running \proglang{R}. 
\subsection{Obtaining inferences from $m>1$ data sets}
 \subsubsection{Averaging the $m>1$ data sets before averaging the results}
When analysing multiple synthetic data sets, traditionally the analyst considers each data set separately before later combining inferences. While point estimates are simply averaged, the way in variance estimates are combined depends on the type of synthesis carried out: specifically, whether fully or partially synthetic data sets are generated and whether the Bayesian framework is used. The combining rules also depend on whether an analyst is using the synthetic data to estimate a population parameter $Q$, or an observed data estimate $\hat{Q}$: the former needs to account for the sampling uncertainty in the original data whereas the latter does not. \par Suppose, then, that an analyst wishes to estimate a univariate population parameter $Q$ from $m>1$ synthetic data sets. A point estimate $q^{(l)}$, and its variance estimate $v^{(l)}$, is obtained from each synthetic data set, $l=1, \hdots, m$. Before these estimates are substituted into a set of combining rules, it is common, as an intermediary step, to first calculate the following three quantities \citep{Drechsler2011}:  
\begin{align*} 
\bar{q}_m &=\frac{1}{m}\sum\limits_{l=1}^m q^{(l)}, & b_m &= \frac{1}{(m-1)}\sum\limits_{l=1}^m (q^{(l)}-\bar{q}_m)^2, & 
\bar{v}_m &= \frac{1}{m}\sum\limits_{l=1}^m v^{(l)}, 
\end{align*}  
where $\bar{q}_m$ is the mean estimate, $b_m$ is the `between-synthesis variance', that is, the sample variance of the $m>1$ estimates, and $\bar{v}_m$ is the mean `within-synthesis variance', the mean of the estimates' variance estimates. \par
The quantity $\bar{q}_m$ is an unbiased estimator for $\hat{Q}$, and is so regardless of whether fully or partially synthetic data sets are generated. The variance estimator, however, does depend on the type of synthetic data. When using saturated models as described in Section \ref{sec2}, partially synthetic data sets are generated because the Bayesian posterior predictive distribution is not used to impute missing information. Hence, the following estimator $T_p$ \citep{Reiter2003}, is valid when estimating $\text{Var}(\hat{Q})$, 
\begin{align*} 
T_p& = b_m/m + \bar{v}_m.  
 \end{align*}
The sampling distribution (if frequentist) or posterior distribution (if Bayesian) of $\hat{Q}$ is a $t$-distribution with $\nu_p = (m-1)\left(1+{m\bar{v}_m}/{b_m}\right)^2$ degrees of freedom. Often, $\nu_p$ is large enough for the $t$-distribution to be approximated by a normal distribution. However, when the between-synthesis variability is much larger than the within-synthesis variability, that is, when $b_m$ is much larger than $\bar{v}_m$ - as may happen when large amounts of noise are applied to protect sensitive records - then $\nu_p$ is crucial to obtaining valid inferences. \par As the data sets are ``completely synthesized'', the following estimator $T_s$ \citep{Raab2016} is valid, too, in large samples,
 \begin{align*}
 T_s& = \bar{v}_m \Big( \frac{n_\text{syn}}{n} + \frac{1}{m} \Big) \approx\bar{v}_m \Big( 1 + \frac{1}{m} \Big).
 \end{align*} 
The large sample assumption facilitates the use of a normal distribution for the sampling distribution (or the posterior distribution) of $\hat{Q}$ when $T_s$ is used to estimate the variance. The overriding advantage of $T_s$ is that, assuming its conditions hold, it allows valid variance estimates to be obtained from $m=1$ synthetic data set.  \par 
When using count models, as opposed to multinomial models, the synthetic data sample size $n_\text{syn}$ is stochastic with mean equal to $n$ and a variance that depends on the synthesis model's tuning parameters, in addition to the distribution of cell sizes in the original data ($\tau_2$). The estimator $T_p$ assumes, implicitly, that $n_\text{syn} = n$ in each of the $m$ synthetic data sets; and $T_s$ assumes that $n_\text{syn}$ is constant across the $m$ synthetic data sets, though not necessarily equal to $n$. When saturated models are used, therefore, and $n_\text{syn}$ is not constant, this assumption is violated which, in turn, may invalidate $T_p$ and $T_s$. However, in a simulation study unreported in this paper, it was found that $T_p$ and $T_s$ do still provide valid variance estimates, that is, confidence intervals achieve the nominal coverage. In some cases, such as when $K$ (the number of cells) is small and $n$ is large, the difference between $n_\text{syn}$ and $n$ may be too large to be ignorable, and then new estimators would be required. Such estimators may introduce weights $w_1\hdots, w_m$ that are related to $n_\text{syn}^{(1)}, \hdots, n_\text{syn}^{(m)}$, the sample sizes of the $m$ synthetic data sets. 
 \subsubsection{Averaging the $m>1$ data sets before analysing them}
 When faced with multiple {categorical} data sets, analysts (and attackers) may either pool or average the data sets \textit{before} analysing them. This is feasible only with contingency tables, as they have the same structure across the $m>1$ data sets. There are several advantages to doing so. Firstly, it means that analysts only have to undertake their analyses once rather than multiple times, thus leading to reduced computational time. Note, although averaging results in non-integer ``counts'', standard software such as the \code{glm} function in \proglang{R} can typically cope with this and still allow models to be fit. Secondly, model-fitting in aggregated data is often hampered by the presence of zero counts, but either averaging or pooling reduces the proportion of zero counts, since it only takes one non-zero across the $m>1$ data sets to produce a non-zero when averaged or pooled. \par When the NBI is used, given an original count $f_i=N$ ($k=1,\hdots,K$), the corresponding mean synthetic cell count $\bar{f}_i^\text{syn}$ has mean and variance given as, 
 \begin{align}
    E(\bar{f}_i^\text{syn})&=N \quad \text{and} \quad \text{Var}(\bar{f}_i^\text{syn})=\frac{1}{m} \big(N+\sigma N^2 \big), \label{meanvar} 
\end{align}
as the synthetic data sets are independent. Thus the effect of using $m>1$ is that the variance of the synthetic cell counts is multiplied by $1/m$.
\section{Introducing the $\tau_3(k,d)$ and $\tau_4(k,d)$ metrics}  \label{sec4}
When multiple synthetic data sets are generated and the mean synthetic count calculated - which is no longer always an integer - it becomes more suitable to consider the proportion of synthetic counts \textit{within a certain distance of} original counts of $k$. To allow this, the metrics $\tau_3(k)$ and $\tau_4(k)$ can be extended to $\tau_3(k,d)$ and $\tau_4(k,d)$, respectively. Mathematically, for an arbitrary cell in the original data $f$ and its corresponding cell in the synthetic data $f^{\text{syn}}$:
\begin{align*}
\tau_3(k,d)&\coloneqq p(|f^{\text{syn}}-k | \leq d  \mid f=k), \\ 
\tau_4(k,d)&\coloneqq p(f=k  \mid |f^{\text{syn}}-k | \leq d ).  
\end{align*}
The metric $\tau_3(k,d)$ is the probability that a cell count of size $k$ in the original data is synthesized to within $d$ of $k$; and The metric $\tau_4(k,d)$ is the probability that a cell count within $d$ of $k$ in the synthetic data originated from a cell of $k$. Unlike $k$, $d>0$ does not need to be an integer. \par
By extending the $\tau_1(k,d)$ metric in a similar way, such that $\tau_1(k,d)$ is the proportion of cell counts in the synthetic data within $d$ of $k$, it follows that $\tau_3(k,d) \tau_2(k) = \tau_4(k,d) \tau_1(k,d)$. \par It is straightforward to see how $\tau_3(k)$ and $\tau_4(k)$ are, in effect, special cases of $\tau_3(k,d)$ and $\tau_4(k,d)$, respectively; the former are returned by setting $d=0$. For small $k$, these $\tau(k,d)$ metrics are intended primarily as risk metrics, because they are dealing with uniques or near uniques. However, when $d$ is reasonably large, $\tau_3(k,d)$ and $\tau_4(k,d)$ are, perhaps, better viewed as utility metrics, because they are dealing with the proportion of uniques that are synthesized to much larger counts (which impacts utility).  \par
For some distributions there are formulae which give exact distributions for sums of random variables, for example, the sum of two Poisson distributions with means $\lambda_1$ and $\lambda_2$ is Poisson distributed with mean $\lambda_1 + \lambda_2$. Alas, for most count distributions, including the NBI, such relationships are less clear. When $m>1$ is sufficiently large, though, tractable expressions for the $\tau_3(k,d)$ and $\tau_4(k,d)$ metrics can be obtained via the Central Limit Theorem (CLT), as the distributions of the mean synthetic cell counts can be approximated by a normal distribution, with mean and variance as given in (\ref{meanvar}). That is, given an original count $f_i=N$ ($k=1,\hdots,K$), when $m$ is large, the distribution of the corresponding mean synthetic cell count $\bar{f}_i^\text{syn}$ is given as:
\begin{align}
    \bar{f}_i^\text{syn}& \mid f_i=N, \sigma, m \sim \text{Normal}(N, (N+\sigma N^2)/m ).\nonumber 
    \intertext{This can be used to approximate $\tau_3(k,d)$ and $\tau_4(k,d)$:}
\tau_3(k,d)&=p(|\bar{f}^{\text{syn}}-k | \leq d \mid f=k), \nonumber \\
&= p(\bar{f}^{\text{syn}}<k+d\mid f=k) - p(\bar{f}^{\text{syn}}<k-d\mid f=k), \nonumber \\
&=\Phi \left(\frac {(k+d)-k }{\sqrt{(k+\sigma k^2)/m} }\right)-\Phi \left(\frac {(k-d)-k }{\sqrt{(k+\sigma k^2)/m} }\right) \nonumber\\
&=2\Phi \left(\frac {d }{\sqrt{(k+\sigma k^2)/m} }\right)-1 \label{tau3}, \\
\tau_4(k,d)&=p(f=k \mid |\bar{f}^{\text{syn}}-k | \leq d ) \nonumber \\
&=  \frac{\tau_3(k,d) \cdot \tau_2(k)}{\sum_{i=0}^{\infty} p(|f^{\text{syn}}-k | \leq d  \mid f=i) \cdot p(f=i) } \nonumber  \\
&= \medmath{\frac{\left[2\Phi \left( {d }\Big/{\sqrt{(k+\sigma k^2)/m} }\right)-1 \right] \cdot \tau_2(k)}{\sum\limits_{i=1}^{\infty} \left[ \Phi \left( {(k+d-i) }\Big/{\sqrt{(i+\sigma i^2)/m} }\right)-\Phi \left( {(k-d-i) }\Big/{\sqrt{(i+\sigma i^2)/m} }\right)\right]\cdot \tau_2(i)}} \label{tau4}
\end{align}
where $\Phi$ is which is used to denote the cumulative distribution function (CDF) of the standard normal distribution.
\section{Empirical Study} \label{sec5}
  The data set synthesized here was constructed with the intention of being used as a substitute to the English School Census, an administrative database held by the Department for Education (DfE). It was generated using public 2011 census output tables involving local authority, sex, age, language and ethnicity; a more detailed description of its origin is available \citep{Jackson2021}. The data itself, however, has little importance here, as the objective is merely to demonstrate the risk and utility of synthetic data and, specifically, the role that $m$ plays in relation to the risk-utility trade-off; hence any categorical data set would suffice. \par The data comprises 8.2 $\times 10^6$ individuals observed over $p=5$ categorical variables. The local authority variable has the greatest number of categories with 326; while sex has the fewest with 4. When aggregated, the resulting contingency table has $K=3.5 \times 10^6$ cells, 90$\%$ of which are unobserved, that is, have a count of zero. The proportion of cells in the original data of size $k$, denoted by $\tau_2(k)$, are given in Table \ref{datasummary2}. \par
  \begin{table}
\caption{\label{datasummary2}The values of $\tau_2(k)$, that is, the empirical proportion of cells in the original data of size $k$.}
\centering
\setlength{\tabcolsep}{7.5pt}
\begin{tabular}{*{8}{c}}
Cell size ($k$) & 0 & 1 & 2 & 3 & 4 & 5 & $ \geq6$ \\ \hline 
$\tau_2(k)$ & 0.9038 & 0.0346 & 0.0148 & 0.0075 & 0.0056 & 0.0038 & 0.0300   
\end{tabular}
\end{table}
 The function \code{rNBI} from the \proglang{R} package \pkg{gamlss.dist} \citep{Stasinopoulos2007} was used to generate multiple synthetic data sets using the $(\sigma, \alpha)$-synthesis mechanism described in Section \ref{sec2}. This was done for a range of $\sigma$, 0, 0.1, 0.5, 2 and 10, and $50$ synthetic data sets were generated for each. This allowed comparisons to be drawn for a range of $m$, for example, taking the first five data sets gives $m=5$, taking the first ten gives $m=10$, etc.
 \subsection{Measuring risk}
Evaluating risk in synthetic data, particularly in synthetic categorical data, is
not always straightforward. Attempting to estimate the risk of re-identification \citep{Reiter2009} is not possible, because the ability to link records is lost when a microdata
set is aggregated, synthesized and disaggregated back to microdata again. More-
over, the correct attribution probability (CAP) risk metric \citep{Taub2018}, which seeks to
quantify the risk of attribute disclosure, relies on the presence of an obvious
sensitive variable, for which there is no obvious candidate in this data set. \par
\begin{figure}
\centering
\includegraphics[height = 4cm]{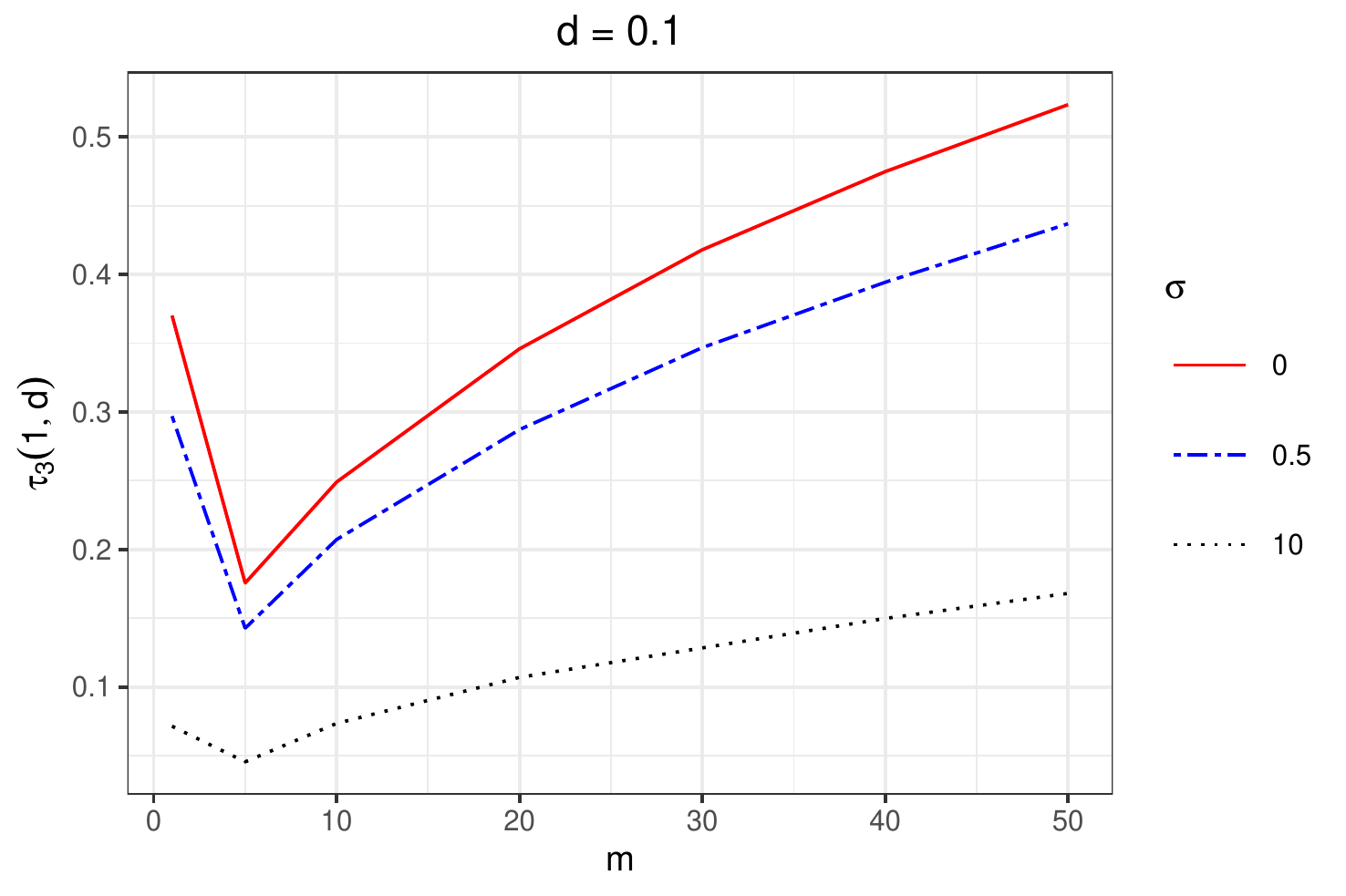}
\includegraphics[height = 4cm]{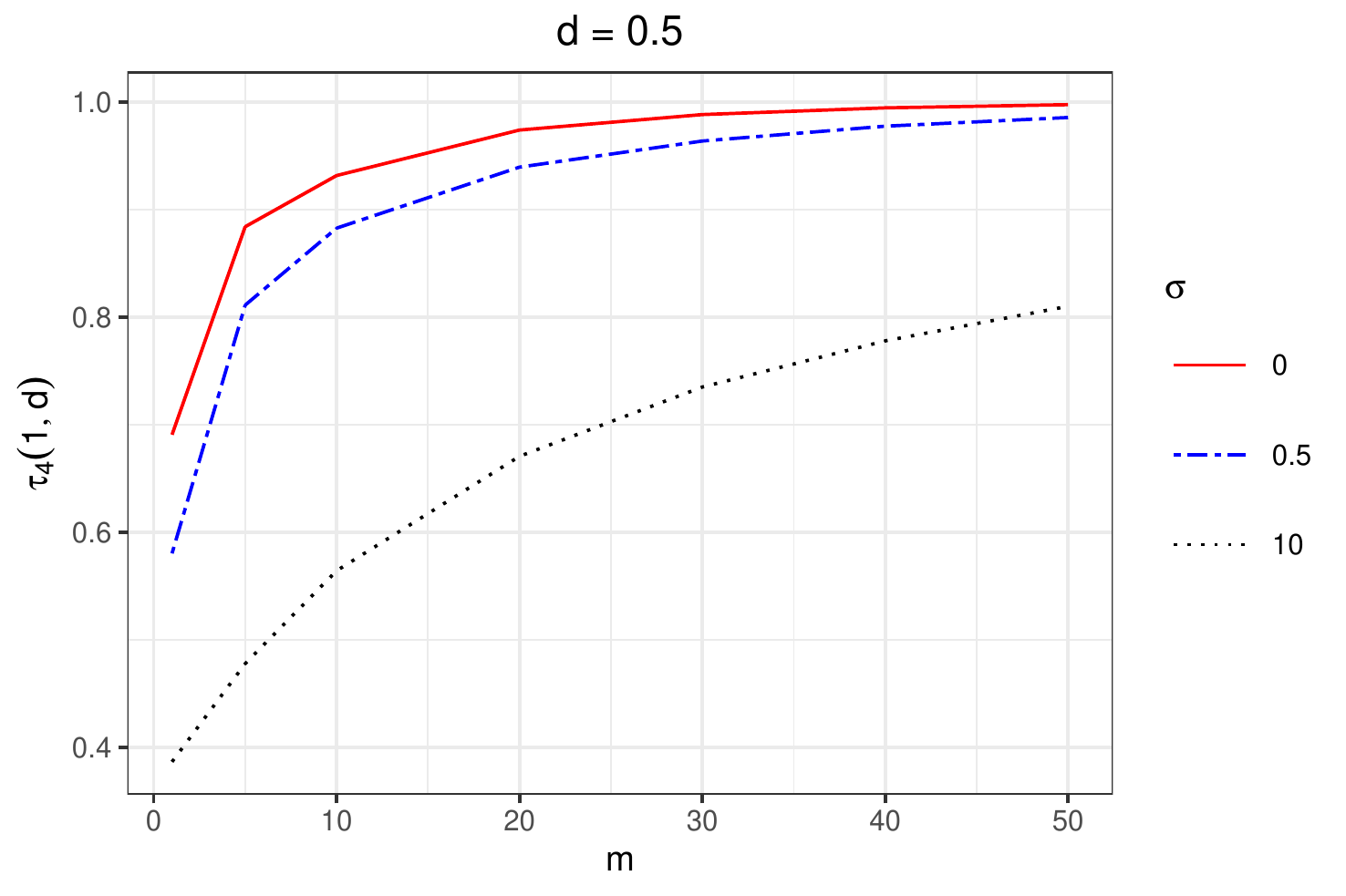}
\includegraphics[height = 4cm]{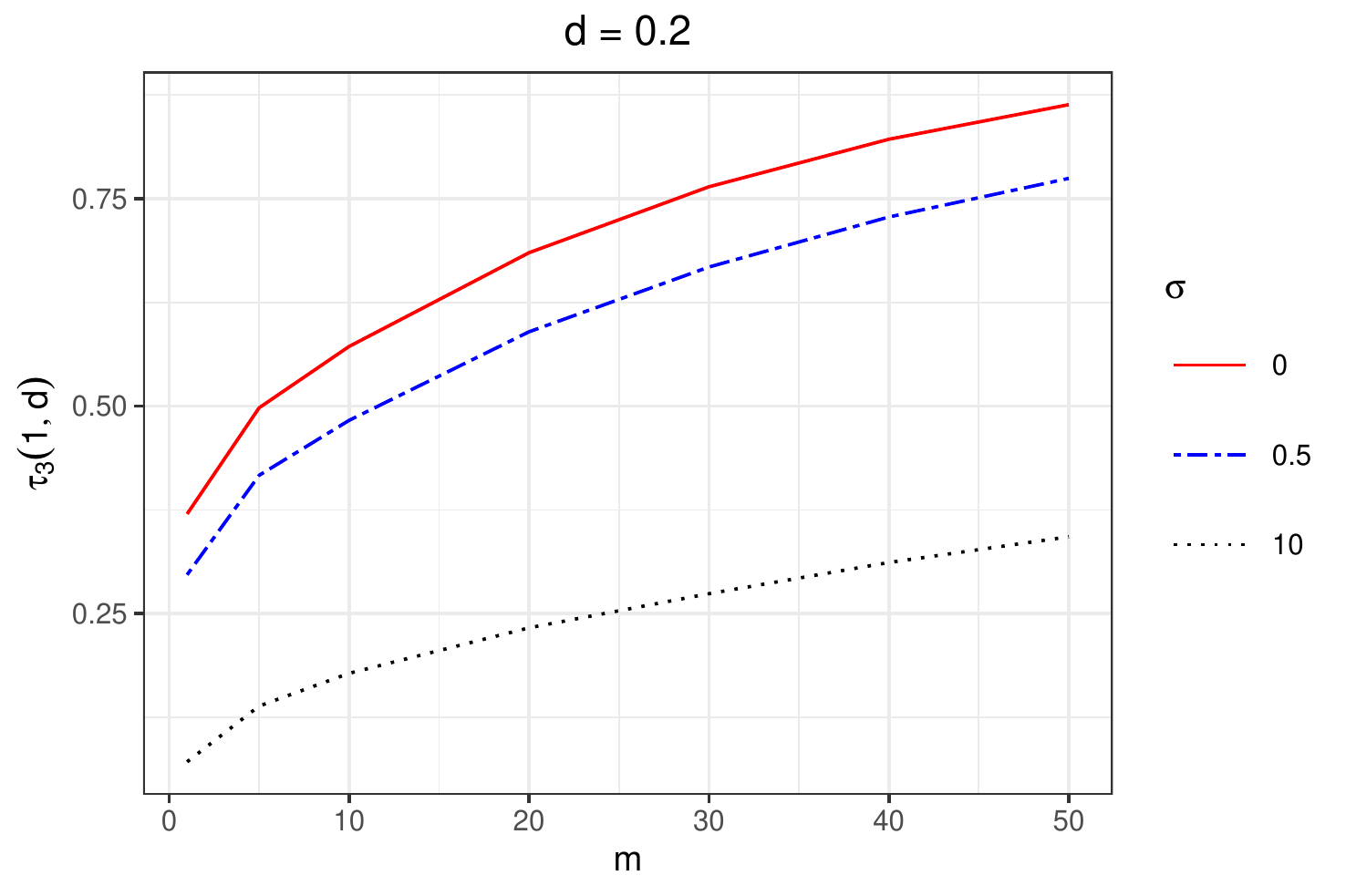}
\includegraphics[height = 4cm]{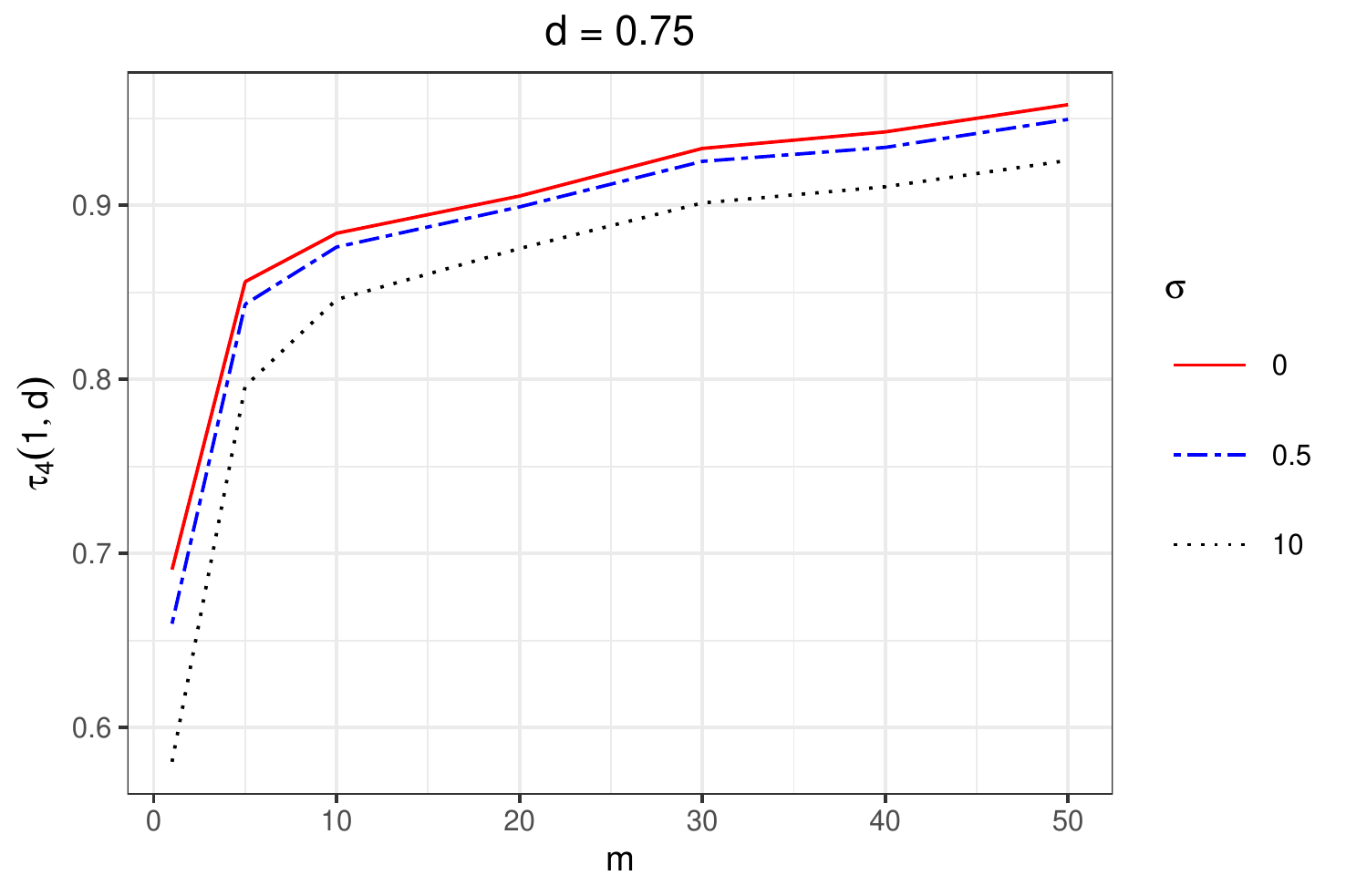}
\caption{\label{fig04}  The left hand plots give the empirical values of $\tau_3(1,d)$ for $d=0.1$ and 0.2; the right hand plots give the empirical values of $\tau_4(1,d)$ for $d=0.5$ and 0.75.}
\end{figure}

Instead, the $\tau_3(1,d)$ and $\tau_4(1,d)$ metrics (that is, setting $k=1$), introduced in Section \ref{sec4}, were used as risk metrics. Figure \ref{fig04} shows that either increasing $m$ or decreasing $\sigma$ increases $\tau_3(1,d)$ and $\tau_4(1,d)$ and hence risk. There is an initial fall in the $\tau_3(1,0.1)$ curves as $m$ increases initially, suggesting lower not higher risk. However, this is just owing to the small $d$: for example, when $d=0.1$, the only way to obtain a mean synthetic count within 0.1 of $k$ when, say $m=5$, is by obtaining a one in each of the five synthetic data sets, compared to just once when $m=1$. \par When $m$ is large, the $\tau_3(k,d)$ and $\tau_4(k,d)$ metrics can be approximated analytically through (\ref{tau3}), which relies on the CLT. There is uncertainty in both the empirical values (owing to simulation error) and the analytical values (owing to the normal approximation), though the divergences between the empirical and analytical values are small. \par
\begin{figure}
\centering
\includegraphics[height = 5cm]{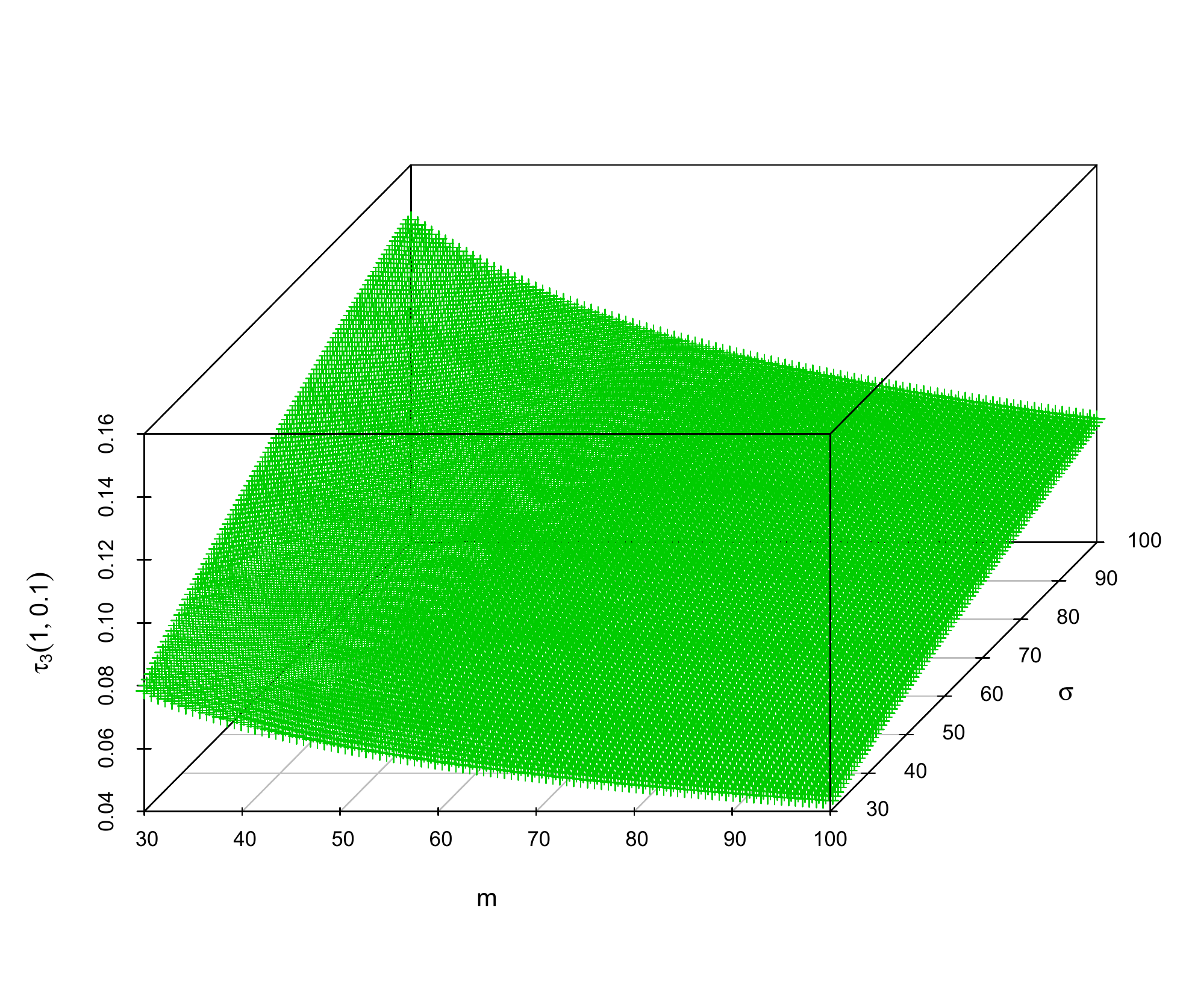}
\caption{\label{fig05} The expected $\tau_3(1,0.1)$ values for $m$ and $\sigma$ greater than 30.}
\end{figure}

 In general, then, increasing $m$ or decreasing $\sigma$ increases risk. This is also shown visually in Figure \ref{fig05}, which demonstrates how $m$ and $\sigma$ can be used in tandem to adjust risk. Here, $\tau_3(1,0.1)$ is used as the $z$-axis (risk) but any $\tau_3(k,d)$ or $\tau_4(k,d)$ would give similar results.  
  \subsection{Measuring utility}
\begin{figure}
\centering
\includegraphics[height = 6cm]{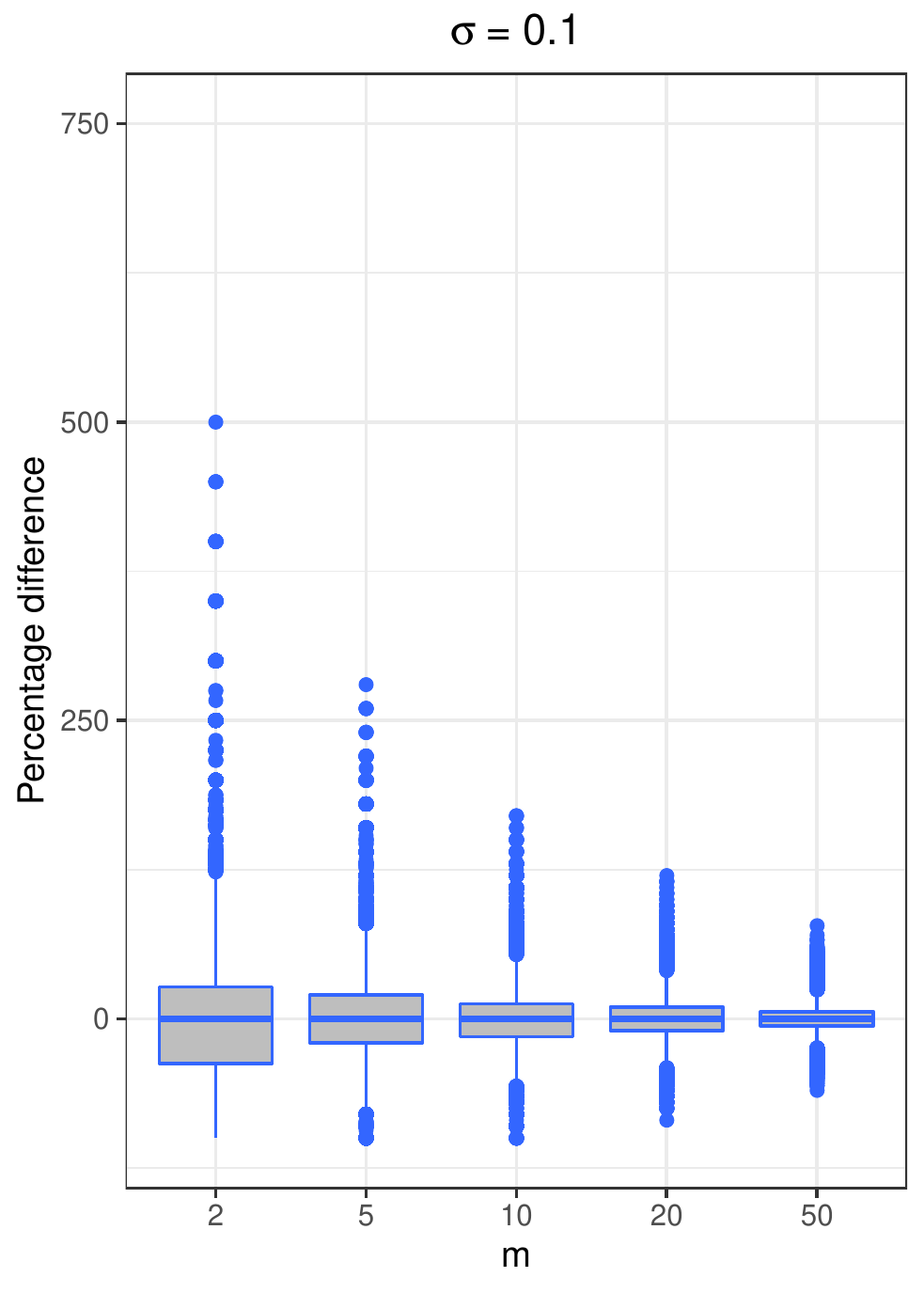}
\includegraphics[height = 6cm]{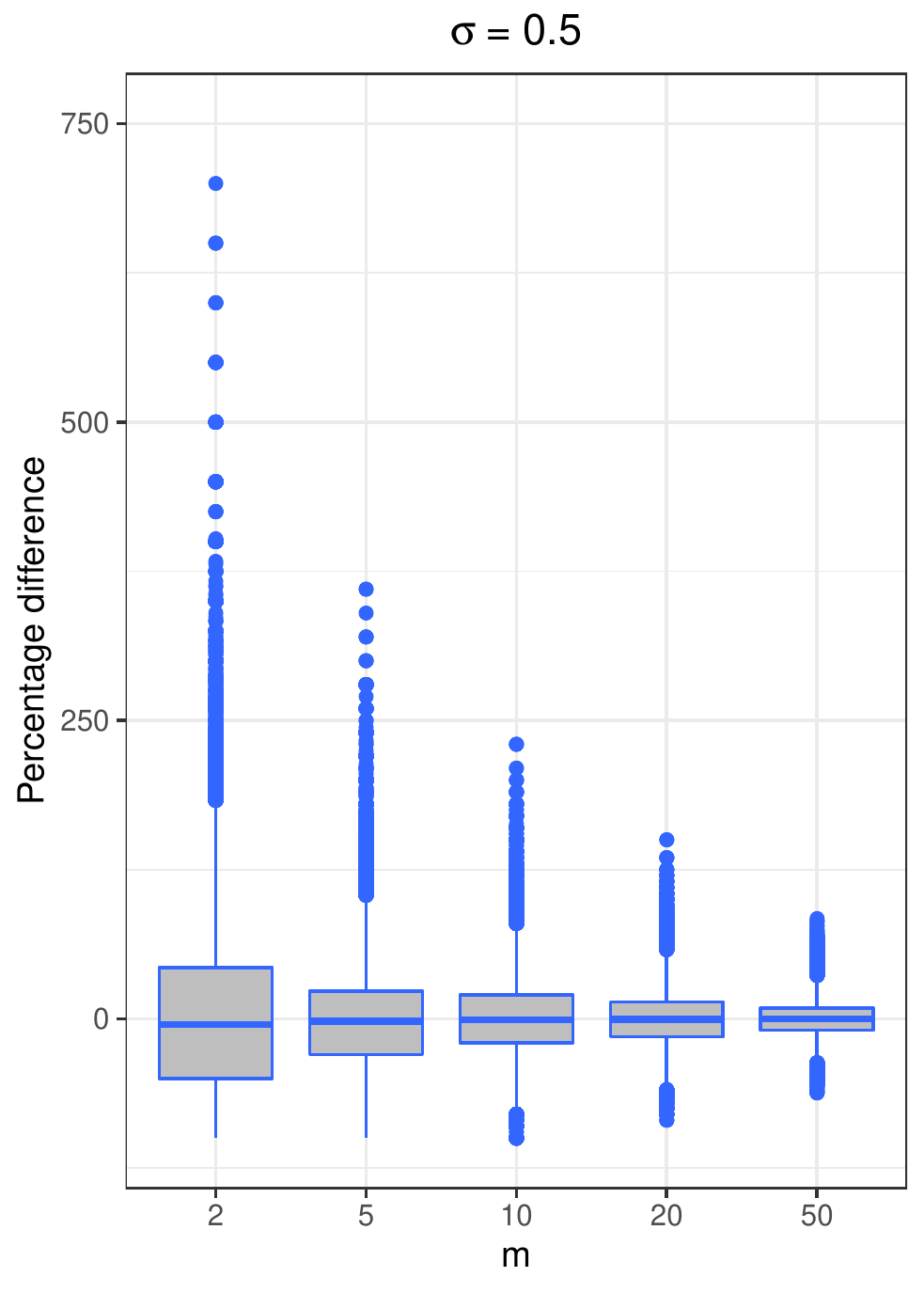}
\caption{\label{fig01} These plots show how increasing $m$ reduces the percentage differences between the mean synthetic counts and the original counts, hence increasing both risk and utility. Both $m$ and $\sigma$ are effective in eliminating the large percentage differences which are present when $m$ is small.}
\end{figure}

\begin{figure}
\centering
\includegraphics[height = 4cm]{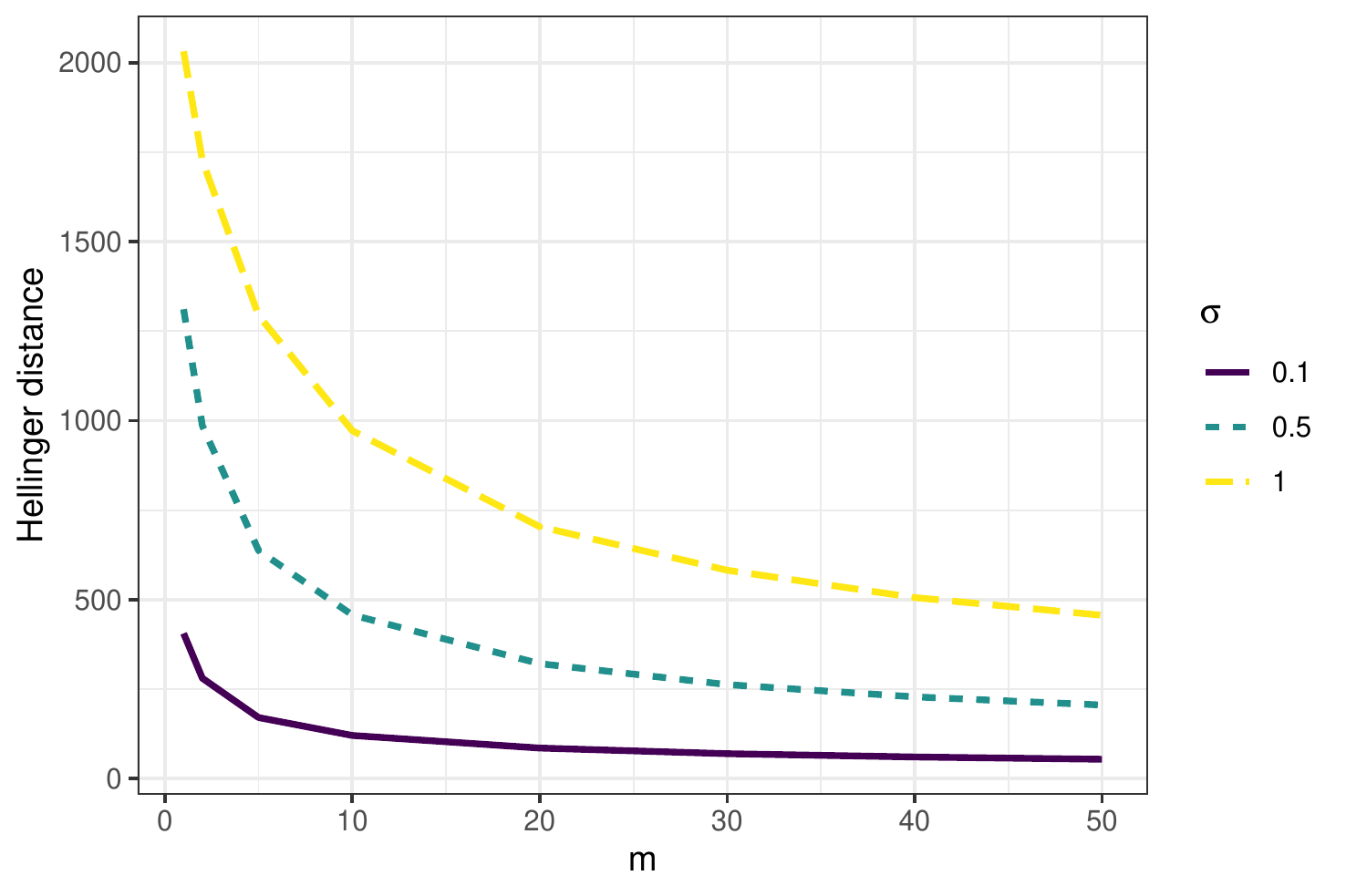}
\includegraphics[height = 4cm]{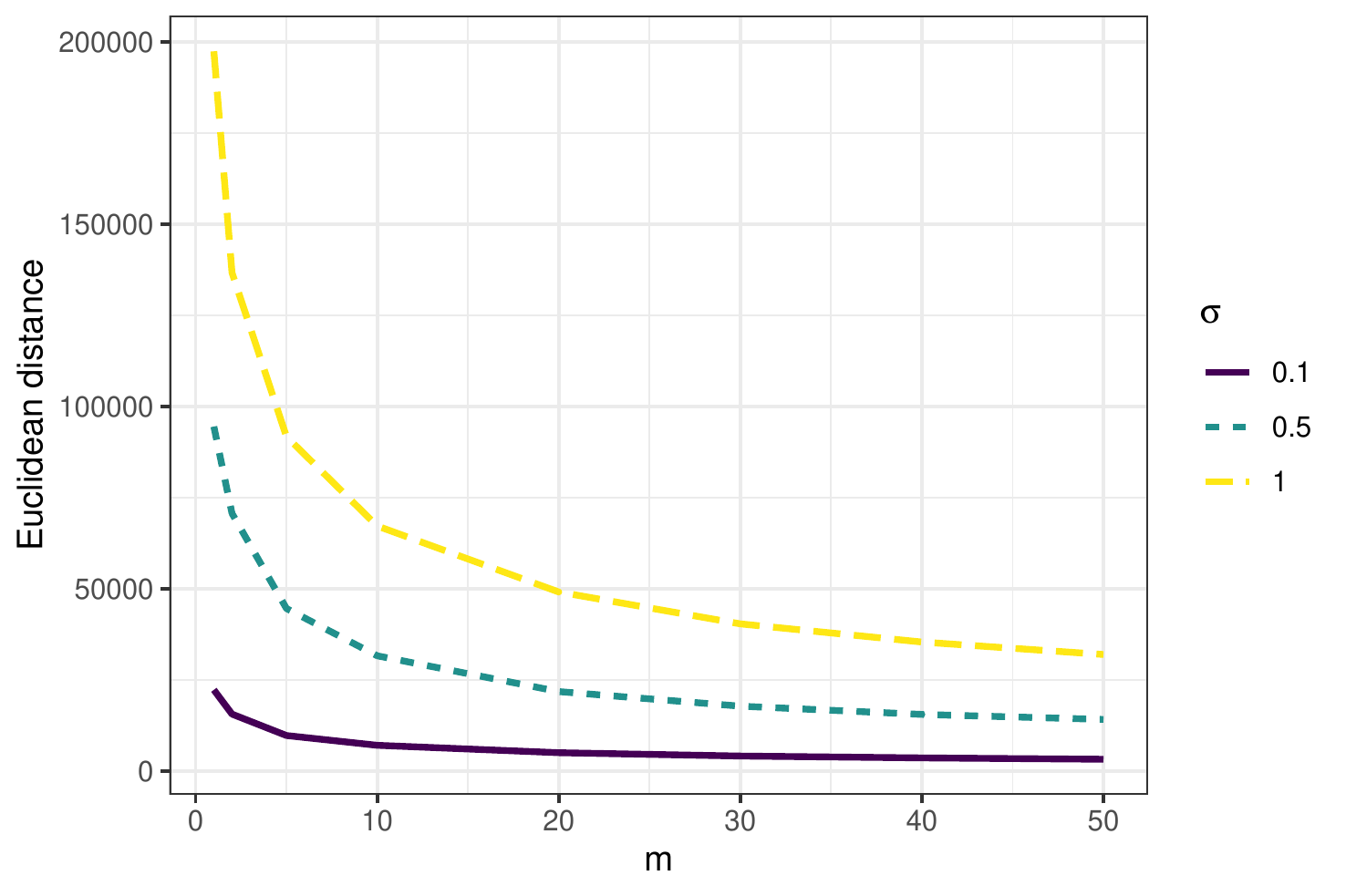}
\caption{\label{fig06} The Hellinger and Euclidean distances as $m$ increases, for various values of $\sigma$. These plots have been created using the cell counts rather than the cell probabilities, though the two are proportional.}
\end{figure}

  As saturated models are used, increasing $m$ (for a given $\sigma$) causes the mean synthetic counts to tend towards the original counts. This can be seen in the boxplots in Figure \ref{fig01}, which show that either increasing $m$ or reducing $\sigma$ reduces the percentage differences between the mean synthetic counts and the original counts. Similarly, the Hellinger and Euclidean distances shown in Figure \ref{fig06} show an improvement in general utility when either increasing $m$ or reducing $\sigma$. \par These measures are equally relevant to risk, too, hence Figures \ref{fig01} and \ref{fig06} reiterate that risk increases with $m$. It is fairly trivial, however, that reducing simulation error increases risk and utility. It is more useful to gain an insight into the \textit{rate} at which risk and utility increase with $m$, that is, the shape of the curves. For example, Figure \ref{fig06}, shows that increasing $m$ has greater effect when $\sigma=1$ than when $\sigma=0.1$.  
The utility of synthetic data can also be assessed for specific analyses by, for example, comparing regression coefficient estimates obtained from a model fit to both the observed and synthetic data. While such measures only assess the synthetic data's ability to support a particular
analysis, they nevertheless can be a useful indicator to, for example, the required $m$ needed to attain a satisfactory level of utility. \par Here, the estimand of interest is a regression coefficient $\beta_1$ from the logistic regression of age $Y$ (aged $\leq9$ = 0, $\geq10$ = 1) on language $X$, that is,
\begin{align*}
Y_i &\sim \text{Bernoulli}(\theta_i) \\
\text{log}\left(\frac{\theta_i}{1-\theta_i} \right) &= \beta_0 + \beta_1 I(X_i=1) \quad \text{for} \quad i = 1, \hdots, n\;(\text{or}\;n_\text{syn}).
\end{align*}
A subset of the data were used, as just two of the language variable's seven categories were considered, while the age variable was dichotomised. When estimated from the original data, $\beta_1$ - which is a log marginal odds ratio - was equal to -0.0075 with a 95$\%$ confidence interval of (-0.0151, -0.0001). Note that, in order to estimate this, it was assumed that the original data constituted a simple random sample drawn from a much larger population. It is hugely doubtful whether such as assumption would be reasonable in practice, but the purpose here was just to evaluate the ability of the synthetic data to produce similar conclusions to the original data. \par
The analysis was undertaken in the two ways described in Section \ref{sec3}. That is, firstly, the $m>1$ synthetic data sets were analysed separately and variance estimates were obtained through the estimator $T_p$. Secondly, the $m>1$ synthetic data sets were pooled into one data set prior to the analysis and variance estimates were obtained through the estimator $T_s$. \par
\begin{figure}
\centering
\includegraphics[height = 6cm]{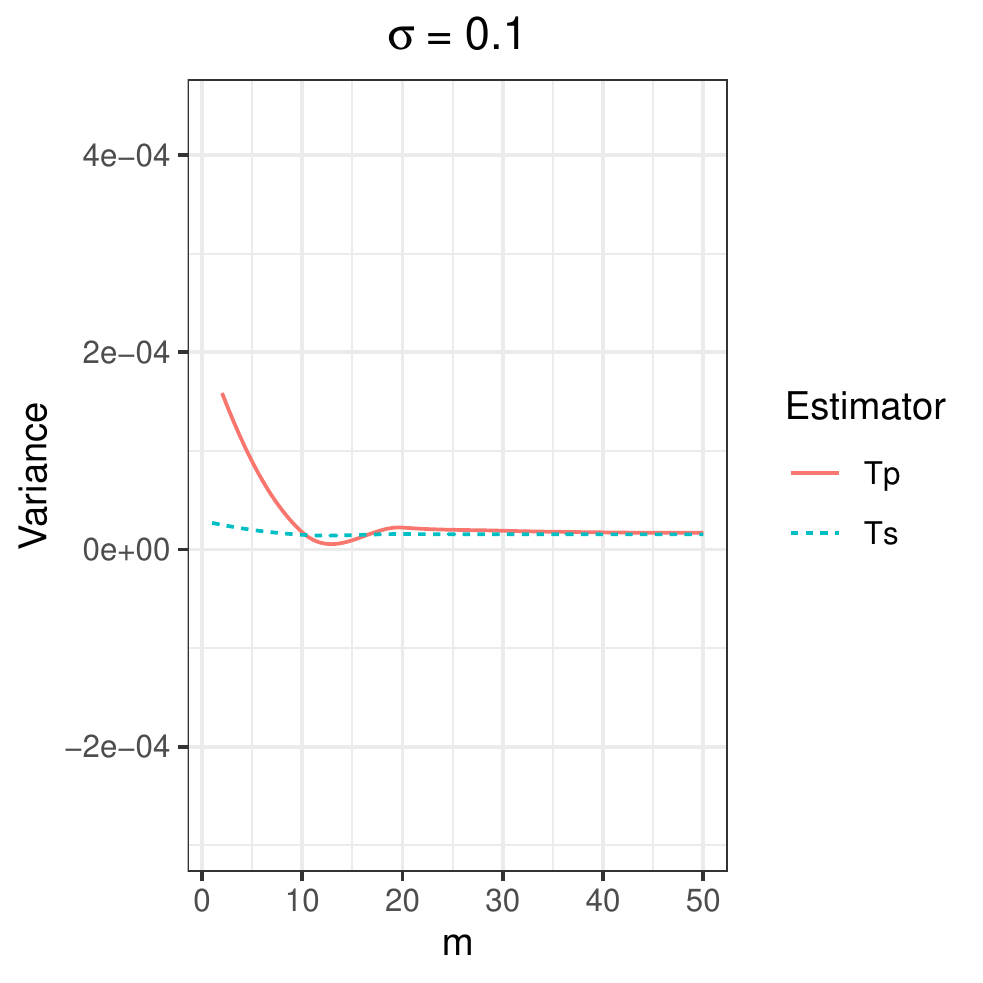}
\includegraphics[height = 6cm]{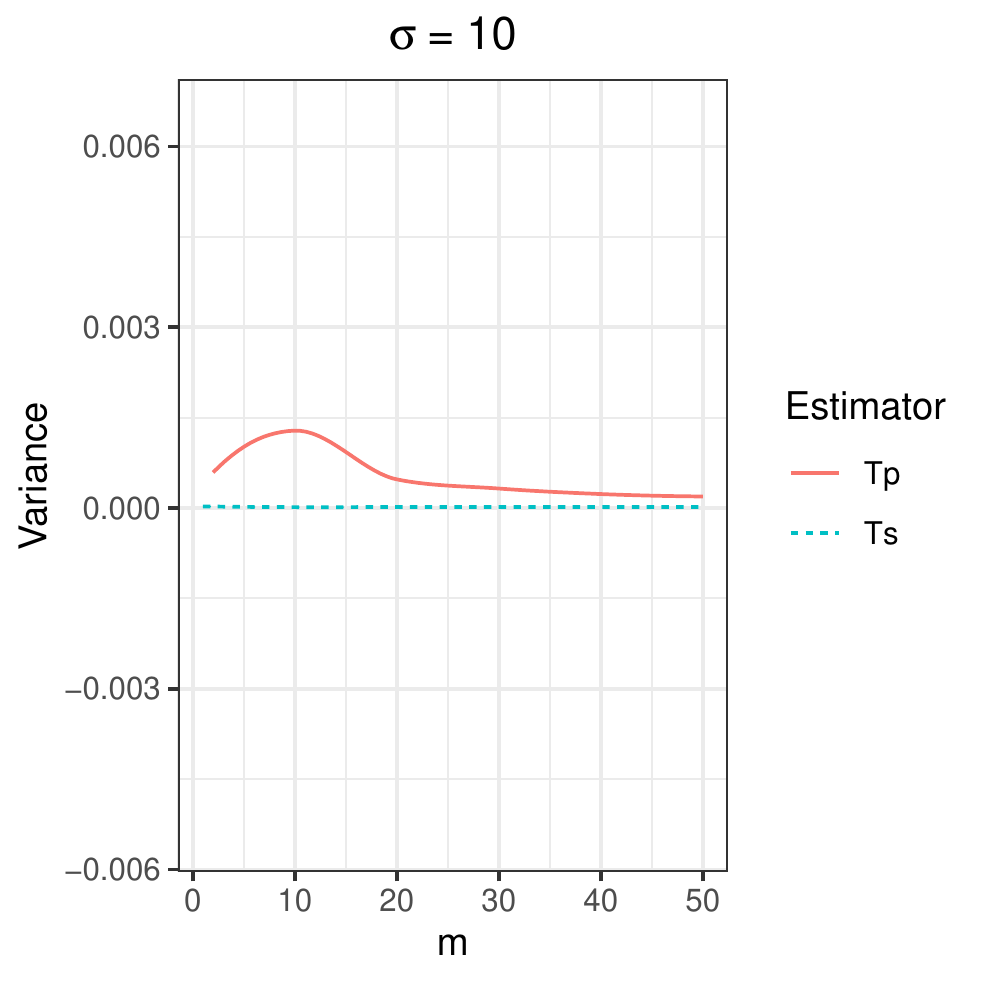}
\caption{\label{fig07} The values of the estimators $T_p$ and $T_s$. For small $m$, $T_p$ is larger than $T_s$, before converging for larger $m$. The estimator $T_s$, on the other hand, remains fairly constant across $m$.}
\end{figure}

 As can be seen in Figure \ref{fig07}, the estimates from $T_p$ were noticeably larger than those from $T_s$, for small $m$. This was worrying for the validity of $T_s$ - and the confidence intervals subsequently computed using $T_s$ - especially since the sampling distribution of $T_p$ was not approximated by a normal distribution, but by a $t$-distribution with $\nu_p$ degrees of freedom, thus widening confidence intervals further. This suggests that the large sample approximation that $T_s$ relies on was not reasonable in this case.      \par
The confidence interval computed from the original data set was compared with the confidence intervals computed from the synthetic data sets via the confidence interval overlap metric \citep{Karr2006,Snoke2018}. This metric is a composite measure that takes into account both the length and the accuracy of the synthetic data confidence interval. Yet whether these factors are weighted appropriately is open to debate. Valid confidence intervals estimated from synthetic data, that is, confidence intervals that achieve the nominal coverage, are longer than the corresponding confidence intervals estimated from the original data, because synthetic data estimates are subject to the uncertainty present in the original data estimates, plus have additional uncertainty from synthesis. However, a synthetic data confidence interval, say, one that is $x\%$ narrower than the original data confidence interval (hence clearly invalid) would yield roughly the same overlap as, say, a confidence interval that is $x\%$ wider. Moreover, either an infinitely wide or infinitely narrow synthetic data confidence interval would achieve an overlap of 0.5. \par The confidence interval overlap results are presented in Table \ref{tab04}. The top frame gives the overlap values from when the data sets are analysed separately, and the bottom frame gives the results from when the data sets are pooled. It can be seen that increasing $m$ broadly results in an increase in the overlap; and that the overlap tends towards 1 as the original and synthetic data confidence intervals converge. The confidence intervals computed using $T_s$ are less robust as those using $T_p$, which is evident in the zero coverage when $m=20$ and $\sigma=10$. This is because, unlike the variance estimator $T_p$, $T_s$ only considers the within-synthesis variability $\bar{v}_m$, not the between-synthesis variability $b_m$. Therefore, if the point estimate deviates, this deviation is not captured by $T_s$.
\begin{table}[p]
\small
\setlength{\tabcolsep}{7.5pt}
\caption{The confidence interval overlap results from when: (i) the data sets were analysed separately and $T_p$ was used to estimate confidence intervals; and (ii) the data sets were pooled and $T_s$ was used to estimate confidence intervals. }
\centering
\begin{tabular}{lcccccccc}
\hline
&& {$m=2$} & {$m=5$} & {$m=10$} & {$m=20$}  & {$m=30$} & {$m=40$} & {$m=50$} \\ 
\multicolumn{2}{c}{}& \multicolumn{7}{c}{}  \\
\hline  \\
\multicolumn{9}{c}{The overlap when the data sets were analysed separately and $T_p$ used}  \\ \\
$\sigma=0$ & & 0.883 & 0.901 & 0.950 & 0.992 & 0.990 & 0.994 & 0.983 \\ 
 $\sigma=0.1$ & & 0.533 & 0.692 & 0.822 & 0.898 & 0.913 & 0.925 & 0.917 \\ 
 $\sigma=0.5$ & & 0.536 & 0.635 & 0.778 & 0.843 & 0.878 & 0.909 & 0.923 \\ 
  $\sigma=2$ & & 0.000 & 0.587 & 0.667 & 0.726 & 0.716 & 0.742 & 0.780 \\ 
  $\sigma=10$ & & 0.522 & 0.535 & 0.554 & 0.583 & 0.604 & 0.623 & 0.638 \\ \\ \\
  \multicolumn{9}{c}{The overlap when the data sets were pooled and $T_s$ used}  \\ \\
$\sigma=0$ & &  0.881 & 0.905 & 0.951 & 0.988 & 0.990 & 0.994 & 0.983 \\ 
  $\sigma=0.1$ &  & 0.700 & 0.317 & 0.802 & 0.942 & 0.904 & 0.920 & 0.915 \\ 
  $\sigma=0.5$ &  & 0.221 & 0.344 & 0.653 & 0.789 & 0.864 & 0.915 & 0.967 \\ 
  $\sigma=2$ &  & 0.020 & 0.436 & 0.856 & 0.775 & 0.825 & 0.809 & 0.906 \\ 
  $\sigma=10$ & & 0.000 & 0.664 & 0.454 & 0.000 & 0.078 & 0.258 & 0.465 \\ 
\end{tabular}
\label{tab04}
\end{table}
\subsection{Tuning $m$ and $\sigma$ in relation to the risk-utility trade-off}
\begin{figure}
\centering
\includegraphics[height = 6cm]{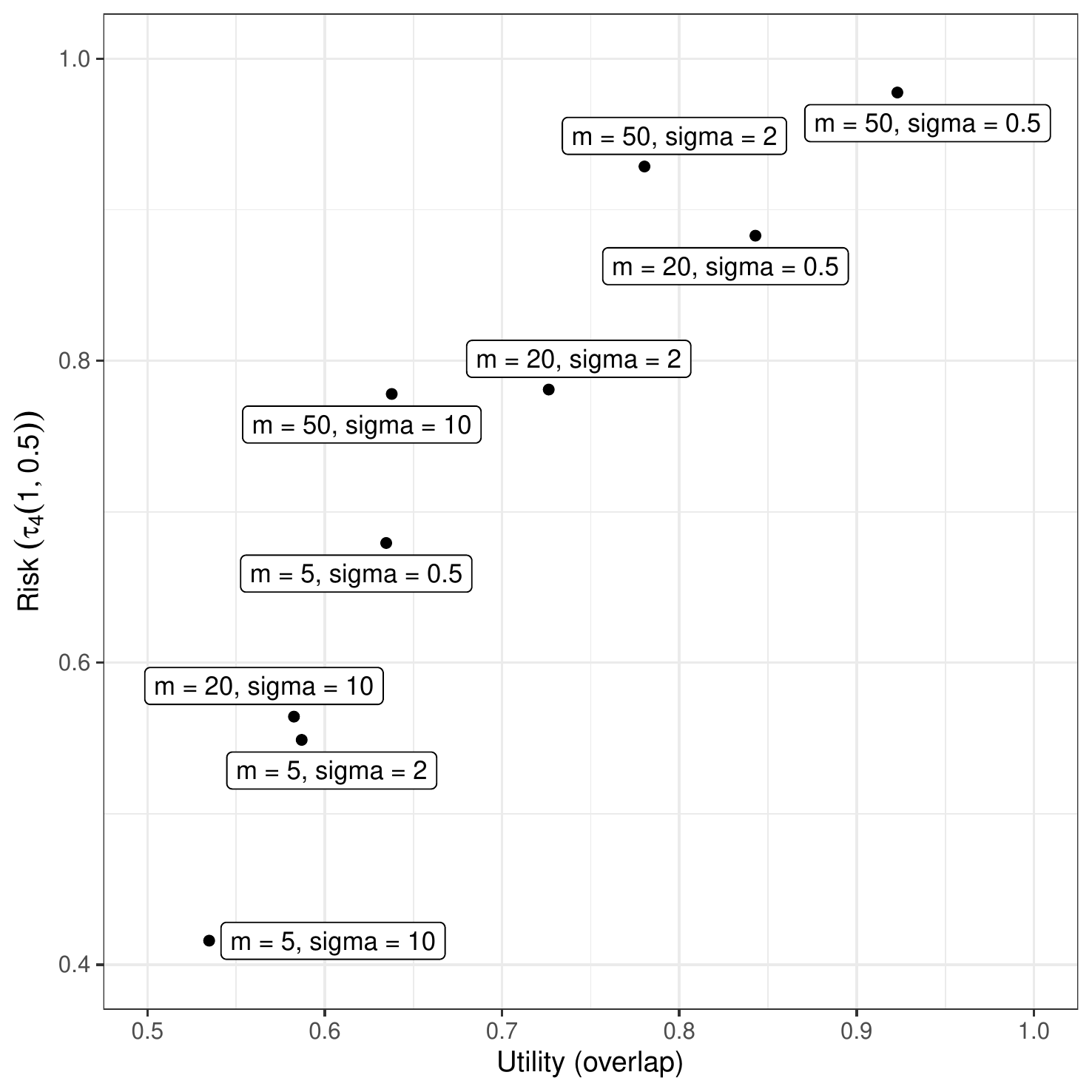}
\includegraphics[height = 6cm]{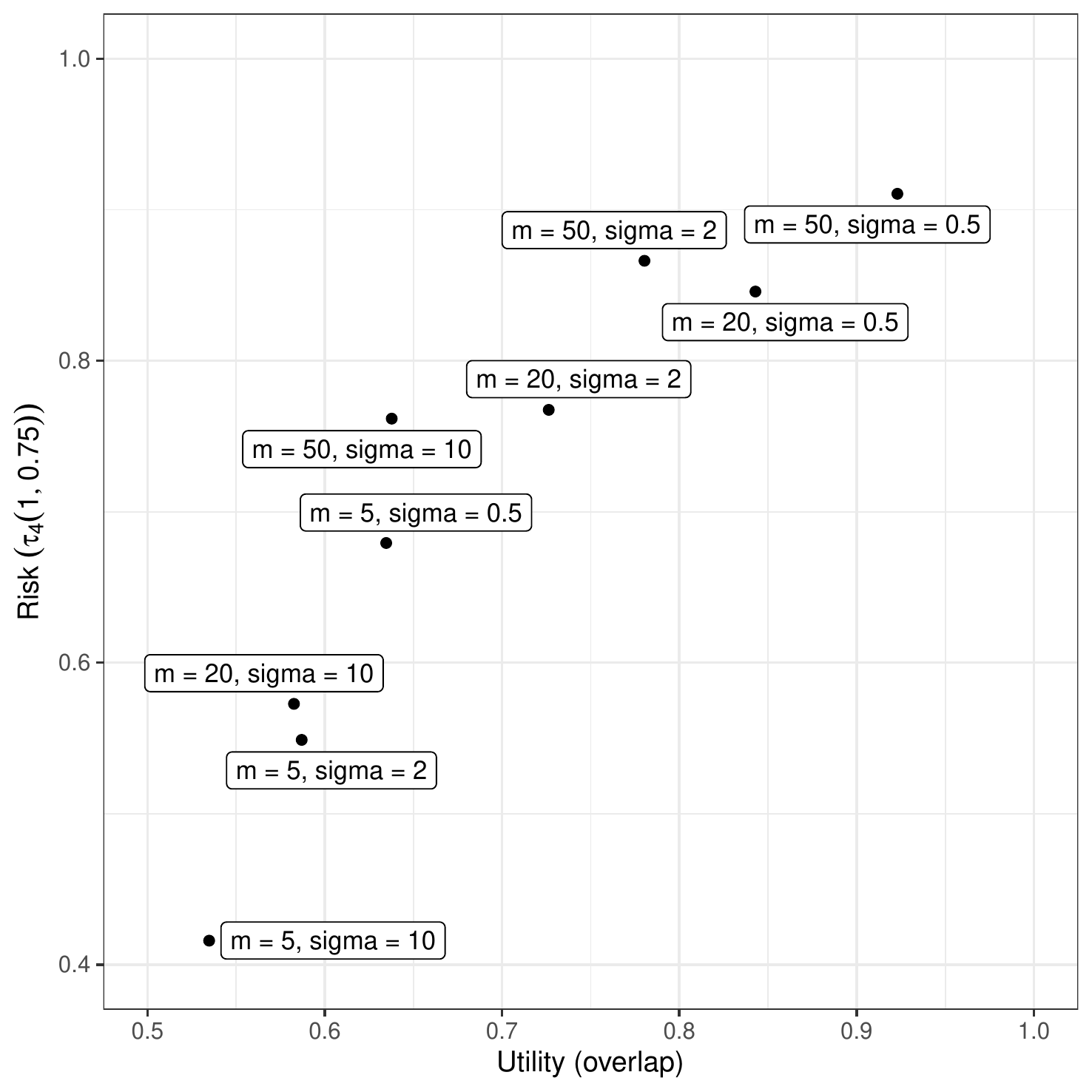}
\includegraphics[height = 6cm]{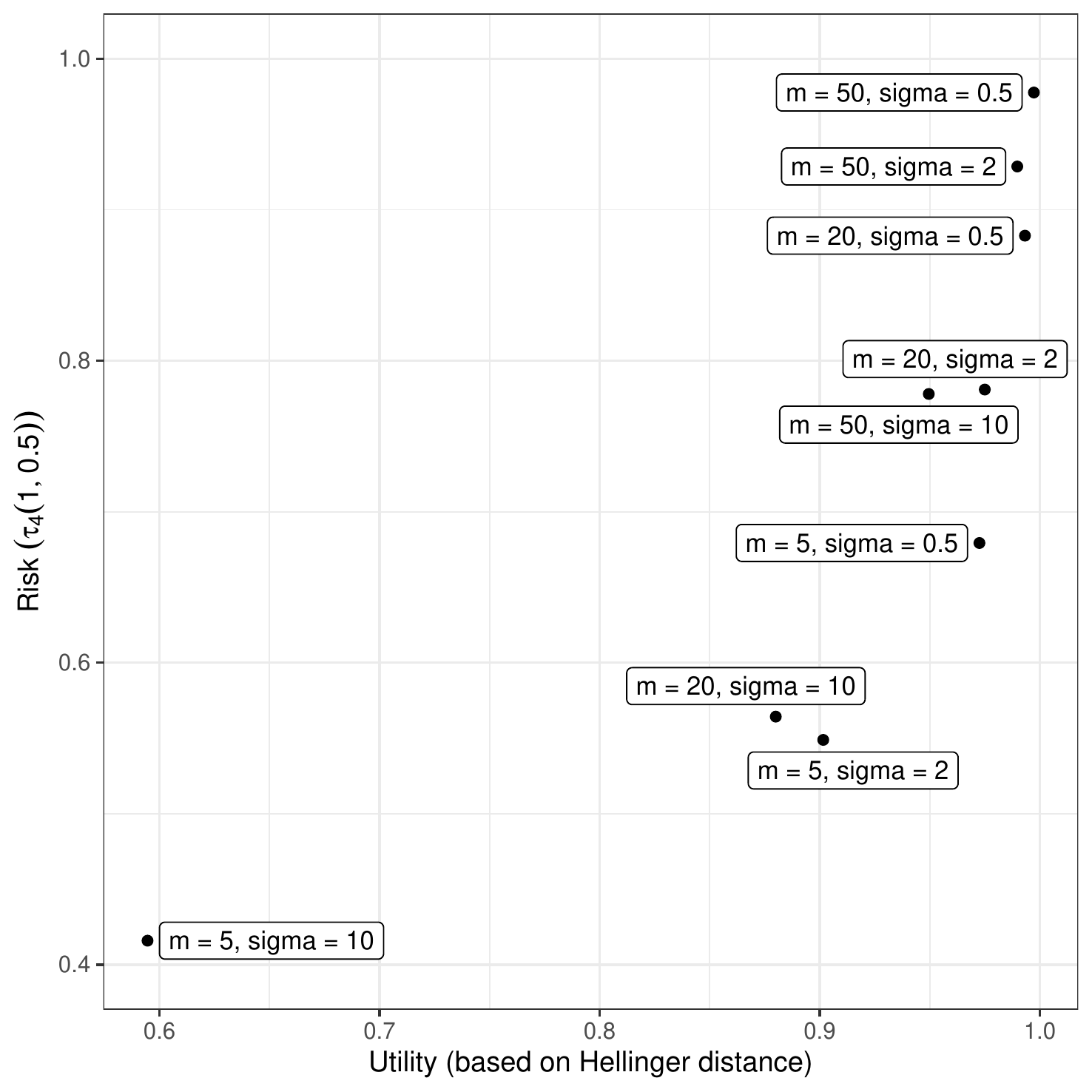}
\includegraphics[height = 6cm]{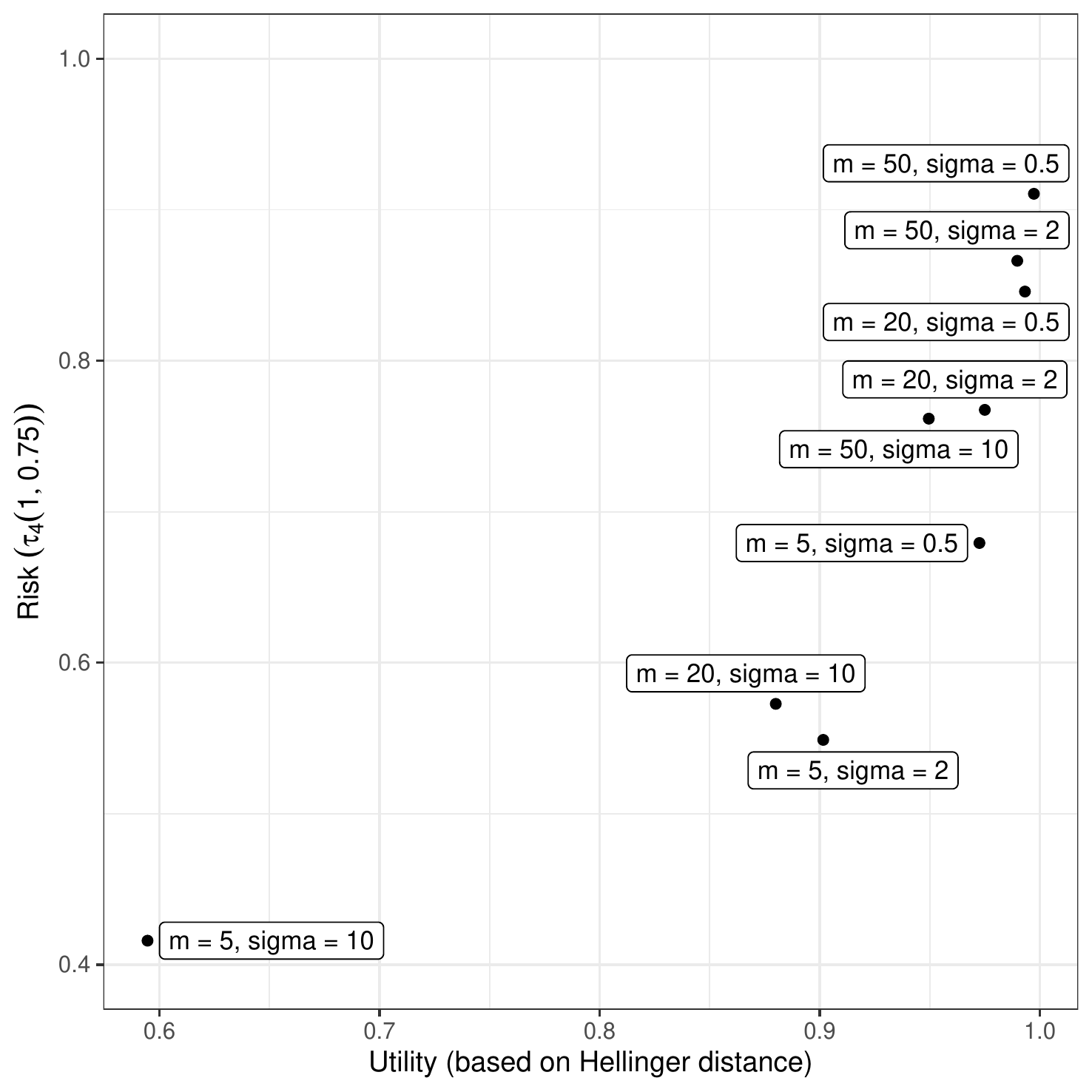}
\caption{\label{fig08} Risk-utility trade-off plots to show where various synthetic data sets are located with respect to the risk-utility trade-off. The optimal position in each plot - that is, the lowest risk and the highest utility - is the bottom right corner. To measure risk, the metrics $\tau_4(1,0.5)$ and $\tau_4(1,0.75)$ were used. To measure utility, the confidence interval overlap and Hellinger distance were used.}
\end{figure}

 The plots in Figure \ref{fig08} show how $m$ and $\sigma$ can be tuned in tandem to produce synthetic data sets that sit favourably within the risk-utility trade-off. These trade-off plots, though, depend on the metrics used to measure risk and utility. Here, risk was measured by either $\tau_4(1,0.5)$ or $\tau_4(1,0.75)$, and utility by either confidence interval overlap (using $T_p$) or Hellinger distance. The Hellinger distances were standardised onto the interval of [0,1] (by dividing by the largest Hellinger distance observed and then were subtracting from 1, so that 1 represents maximum utility and 0 minimum utility). \par Looking at the top-left plot, where risk is measured by $\tau_4(1,0.5)$ and utility measured by confidence interval overlap, synthetic data sets generated with $m=50,\;\sigma=10$ and $m=20,\;\sigma=2$ have roughly the same level of risk. Yet the overlap (utility) for the latter is roughly 0.1 more than the former. It is possible to strictly dominate synthetic data sets over others, that is, obtain lower risk \textit{and} greater utility values. For example, again looking at the top left plot, synthetic data sets generated with $m=50,\;\sigma=2$ have higher risk but lower utility than when $m=20,\;\sigma=0.5$. These visual trade-offs are plotted using the empirical results, so are subject to variation from simulation; the confidence interval overlap values, in particular, can be volatile, especially when $\sigma$ is large. \par The intention is that the synthesizer produces such plots prior to releasing the data. Many metrics, including $\tau_4(k,d)$, can be expressed analytically when using saturated models, which means that the synthesizer does not actually have to generate the synthetic data to produce such plots. 
\section{Discussion}
In addition to $m$, the synthesizer could also  increase or decrease $E(n_\text{syn})$, the expected sample size of each synthetic data set. A single synthetic data set ($m=1$) with ${E}(n_\text{syn})=n$ contains roughly the same number of records as two synthetic data sets ($m=2$) each with ${E}(n_\text{syn})=n/2$. To generate a synthetic data set with an expected sample size of $n/2$, the synthesizer simply takes draws from NBI distributions with means exactly half of what they were previously. Reducing $E(n_\text{syn})$ should reduce risk, as fewer records are released, but inevitably reduces utility, too; once again, it calls for an evaluation with respect to the risk-utility trade-off. \par Moreover, there are further tuning parameters that could be incorporated into this synthesis mechanism. One way would be to use a three-parameter distribution. When using a two-parameter count distribution, the synthesizer can increase the variance but cannot control how the variability manifests itself. The use of three-parameter count distributions would allow the synthesizer to alter the skewness, that is, they could change the shape of the distribution for a given mean and variance. \par 
The estimator $T_s$ \citep{Raab2016} relies on a large sample assumption, thus allowing the use of a normal distribution for the sampling distribution (or the posterior distribution) of $\hat{Q}$. When the sample is not `large', as with the estimator $T_p$, it may be possible to approximate the sampling distribution by the $t$-distribution. A greater understanding of the asymptotic properties would ensure the validity of resulting inferences. An alternative in this instance, that is, when the underlying sampling distribution is unknown, is to use bootstrap methods and calculate, for example, confidence intervals using the `bootstrap-$t$ method' \citep{Efron1993}. \par  This issue with sample size seems to be linked to a wider issue with tabular data analysis that stretches beyond synthetic data. It is often unclear whether sample size in tabular data should refer to $n$ (the number of subjects) or $K$ (the number of cells). This has implications when using, for example, the Bayesian Information Criterion (BIC), thus can affect model selection. Using $K$ appears to be more intuitive, given that it is these counts that are modelled and subsequently simulated from.      \par 
There are, of course, disadvantages to generating $m>1$ synthetic data sets with the most obvious being the increased risk. Nevertheless, the potential benefits warrant further exploration, especially in relation to the risk-utility trade-off: does the gain in utility outweigh the increase in risk? \par Organisations are taking a greater interest in making data - such as administrative data - available to researchers, by producing their own synthetic data. For this to be successful, organisations need to guarantee the protection of individuals' personal data - which, as more data becomes publicly available, becomes ever more challenging - while also producing data that are useful for analysts. Therefore, there needs to be scope to tune the risk and utility of synthetic data effectively, and considering $m$ as a tuning parameter helps to achieve this.
\bibliography{multiple.bib}
\end{document}